\documentclass[12pt]{article}

\usepackage{epsf}
\usepackage{graphicx}
\usepackage{axodraw} 
\usepackage{rotating}
\usepackage{epic}
\usepackage{curves}
\usepackage{epsf}
\usepackage{rotate}

\renewcommand{\thefootnote}{\fnsymbol{footnote}}

\begin{document}

%%%%%%%%%%% Titlepage %%%%%%%%%%% 

\begin{titlepage}
\begin{flushright}
\begin{tabular}{l}
DESY 00--052\\
ZH--TH 7/00\\
hep--ph/0004010\\
April 2000
\end{tabular}
\end{flushright}

\vspace*{1.8truecm}

\begin{center}
\boldmath
{\Large \bf Exploring CP Violation with $B_c$ Decays}
\unboldmath

\vspace*{2.5cm}

{\sc{\large Robert Fleischer}}${}^{1,}$\footnote{E-mail: 
{\tt Robert.Fleischer@desy.de}} \,\, {\large and }\,\,
{\sc{\large Daniel Wyler}}${}^{2,}$\footnote{E-mail: 
{\tt wyler@physik.unizh.ch}}\\[0.8cm]
\vspace*{0.1cm} ${}^1${\it Deutsches Elektronen-Synchrotron DESY, 
Notkestr.\ 85, D--22607 Hamburg, Germany}\\[0.3cm]
\vspace*{0.1cm} ${}^2${\it Institut f\"ur Theoretische Physik, Universit\"at
Z\"urich, CH--8057 Z\"urich, Switzerland}

\vspace{2.4truecm}

{\large\bf Abstract\\[10pt]} \parbox[t]{\textwidth}{
We point out that the pure ``tree'' decays $B_c^\pm\to D^\pm_s D$ are 
particularly well suited to extract the CKM angle $\gamma$ through
amplitude relations. In contrast to conceptually similar strategies 
using $B^\pm\to K^\pm D$ or $B_d\to K^{\ast0} D$ decays, the advantage 
of the $B_c$ approach is that the corresponding triangles have three 
sides of comparable length and do not involve small amplitudes. Decays 
of the type $B_c^\pm\to D^\pm D$ -- the $U$-spin counterparts of 
$B_c^\pm\to D^\pm_s D$ -- can be added to the analysis, as well as 
channels, where the $D^\pm_s$- and $D^\pm$-mesons are replaced by 
higher resonances.
}

\vskip1.5cm

\end{center}

\end{titlepage}

\thispagestyle{empty}
\vbox{}
\newpage
 
\setcounter{page}{1}

\setcounter{footnote}{0}
\renewcommand{\thefootnote}{\arabic{footnote}}

\noindent 
CP violation is one of the least understood aspects of particle physics
\cite{jarl,bs}. The Standard Model provides a simple description of this 
phenomenon through the complex Cabibbo--Kobayashi--Maskawa (CKM) matrix 
\cite{ckm}, which is consistent with present particle physics experiments.
However, the baryon asymmetry of the Universe clearly requires additional
sources for CP violation \cite{shapos}; therefore CP violation might be 
the road to new physics.
  
Decays of $B$-mesons provide a rich ground for investigating CP violation 
\cite{bl,babar}. They allow both for stringent tests of the Standard Model, 
and for studies of new sources for this effect. Consequently, the 
literature in this field grows daily. Within the Standard Model, CP violation 
is often characterized by the so-called unitarity triangle \cite{UT} and the 
measurements of its three angles $\alpha$, $\beta$ and $\gamma$. Many years 
ago, Bigi, Carter and Sanda showed that these angles could be determined 
through ``mixing-induced'' CP asymmetries, which arise in decays of neutral 
$B$-mesons \cite{bisa}. The most prominent decay is 
$B_d\to J/\psi\, K_{\rm S}$, where the angle $\beta$ can be obtained with 
essentially no theoretical uncertainty. Similarly, the angle $\alpha$ could 
be determined from $B_d\to\pi^+\pi^-$. Unfortunately, it was found later 
that this determination is not theoretically clean because of penguin 
contributions, leading to considerable hadronic uncertainties. These
could be overcome by measuring all $B\to\pi\pi$ decays, in particular 
$B_d\to\pi^0\pi^0$ \cite{gl}. However, this mode is extremely difficult -- 
if not impossible -- to measure. Even worse seemed the situation of the 
angle $\gamma$. 

Since then, interesting new methods to extract this angle with few 
theoretical uncertainties were devised. For instance, it was shown 
that $\gamma$ could be determined through the measurement of six 
$B^{\pm}\to K^\pm D$ decay rates \cite{gw}. To this end, the CP eigenstate 
\begin{equation}\label{D-CP}
|D^0_+\rangle=\frac{1}{\sqrt{2}}\left(|D^0\rangle+
|\overline{D^0}\rangle\right)
\end{equation}
of the neutral $D$-meson system is employed, allowing the derivation of
amplitude triangle relations. Unfortunately, the corresponding triangles 
in the complex plane, which are fixed through the magnitudes of the 
$B^{\pm}\to K^\pm D$ decay amplitudes, turned out to be highly stretched,
and are -- from an experimental point of view -- not very useful to 
determine $\gamma$. Further difficulties were pointed out in \cite{ads}.
As an alternative, the decays $B_d\to K^{\ast 0}D$ were proposed \cite{dun} 
because the triangles are more equilateral. But all sides are small 
because of various suppression mechanisms. In another paper, the triangle 
approach to extract $\gamma$ was also extended to the $B_c$ system by Masetti 
\cite{masetti}.

Another road towards the extraction of $\gamma$, which will not be touched 
here, is provided by $SU(3)$ relations between $B\to\pi K$, $\pi\pi$ decay 
amplitudes \cite{BpiK}. Although this approach is not theoretically clean -- 
in contrast to the $B\to KD$ strategies using pure ``tree'' decays -- it 
is more promising from an experimental point of view. In the context of the 
$B\to\pi K$ modes, it was pointed out that non-trivial bounds on $\gamma$ 
could be obtained \cite{FM}. Also here, it was noted later that other decays 
than the original ones may provide more powerful bounds on $\gamma$ \cite{NR}.
Many recent papers review and extend the situation \cite{BpiK-revs}. 

A comment on the implications of these different methods might be in order. 
As the $B\to KD$ triangle approaches rely on pure ``tree'' decays, 
i.e.\ do not involve any flavour-changing neutral-current (FCNC) 
processes, it is expected that they are not affected significantly by
new physcis (unless it affects $D^0$--$\overline{D^0}$ mixing) and probe 
indeed the angle $\gamma$ as defined in the Standard Model. On the 
other hand, the $B\to\pi K$ methods are strongly sensitive 
to penguin, i.e.\ loop diagrams \cite{NP}. Since these can be influenced 
by new physics, the thus determined value of $\gamma$ may be different from 
the Standard-Model expectation. Consequently, a comparison of the values
of $\gamma$ obtained from pure ``tree'' decays and penguin-dominated
modes would be a good way to search for new physics. Moreover, the
values for $\gamma$ could be compared with the usual fits of the unitarity 
triangle \cite{al}.

\begin{figure}
\begin{center}
\leavevmode
\epsfysize=4.3truecm 
\epsffile{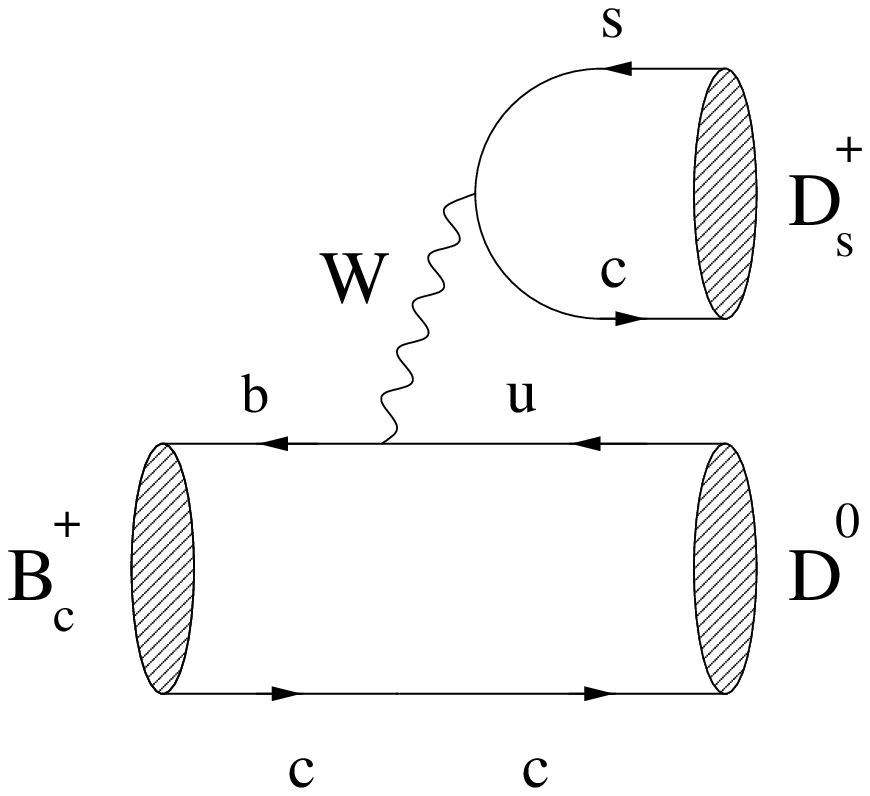} \hspace*{1.8truecm}
\epsfysize=4.7truecm 
\epsffile{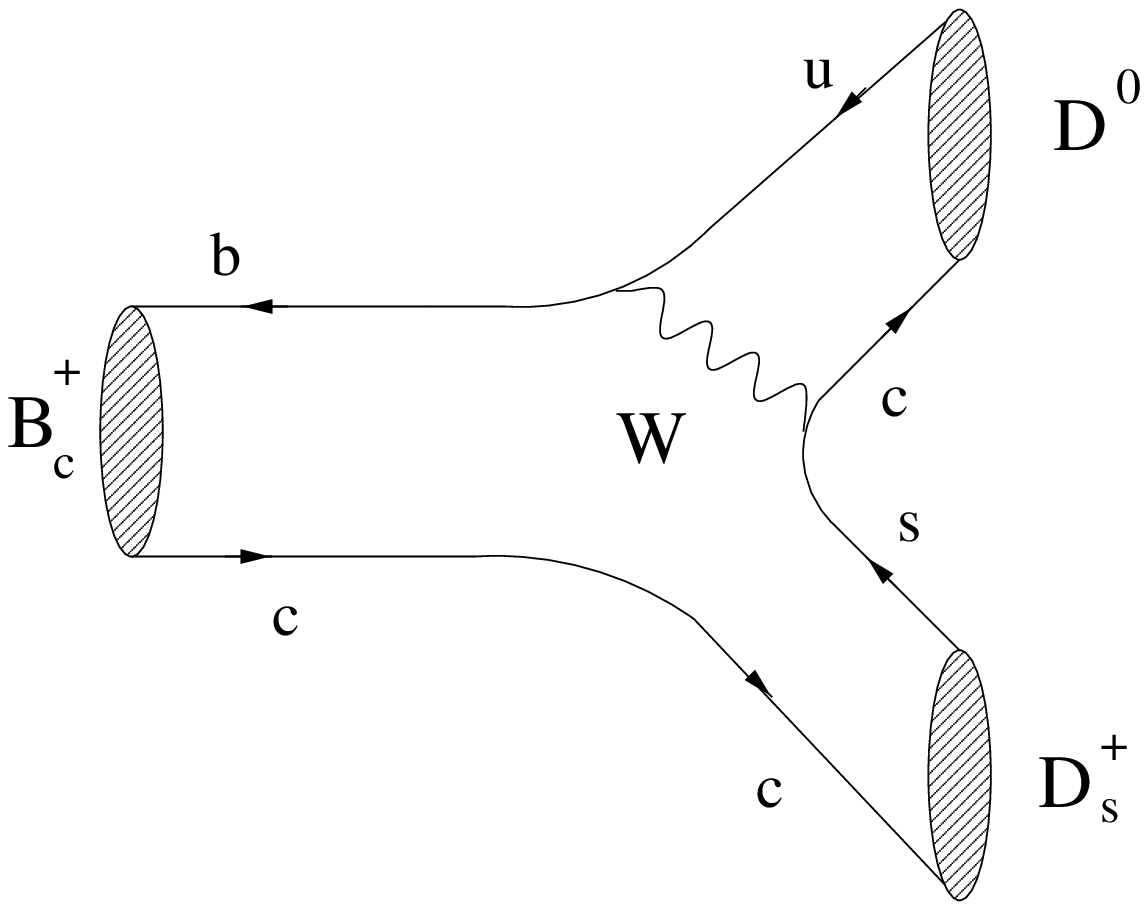}
\end{center}
\vspace*{-0.4truecm}
\caption{Feynman diagrams contributing to the decay 
$B^+_c\to D_s^+D^0$.}\label{fig:BcDD}
\end{figure}

\begin{figure}
\begin{center}
\leavevmode
\epsfysize=4.7truecm 
\epsffile{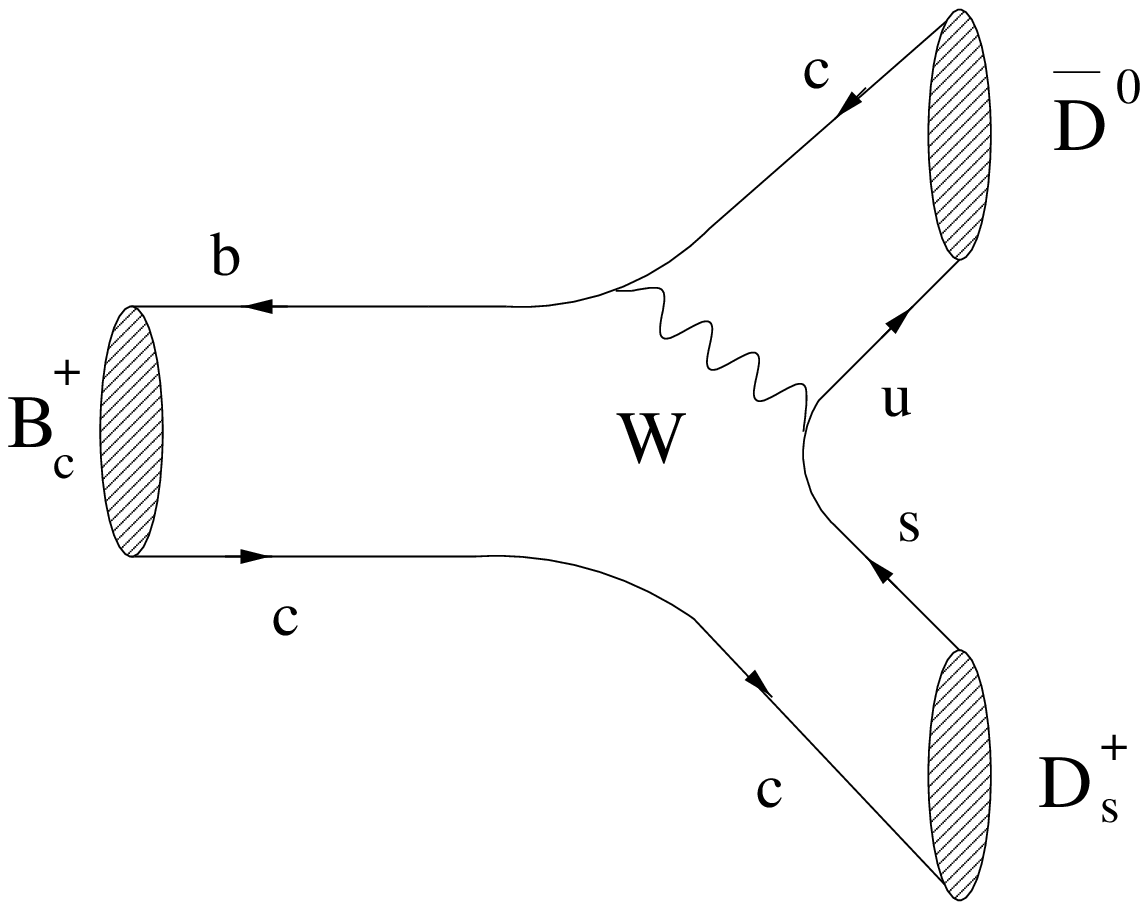} \hspace*{1.8truecm}
\epsfysize=4.5truecm 
\epsffile{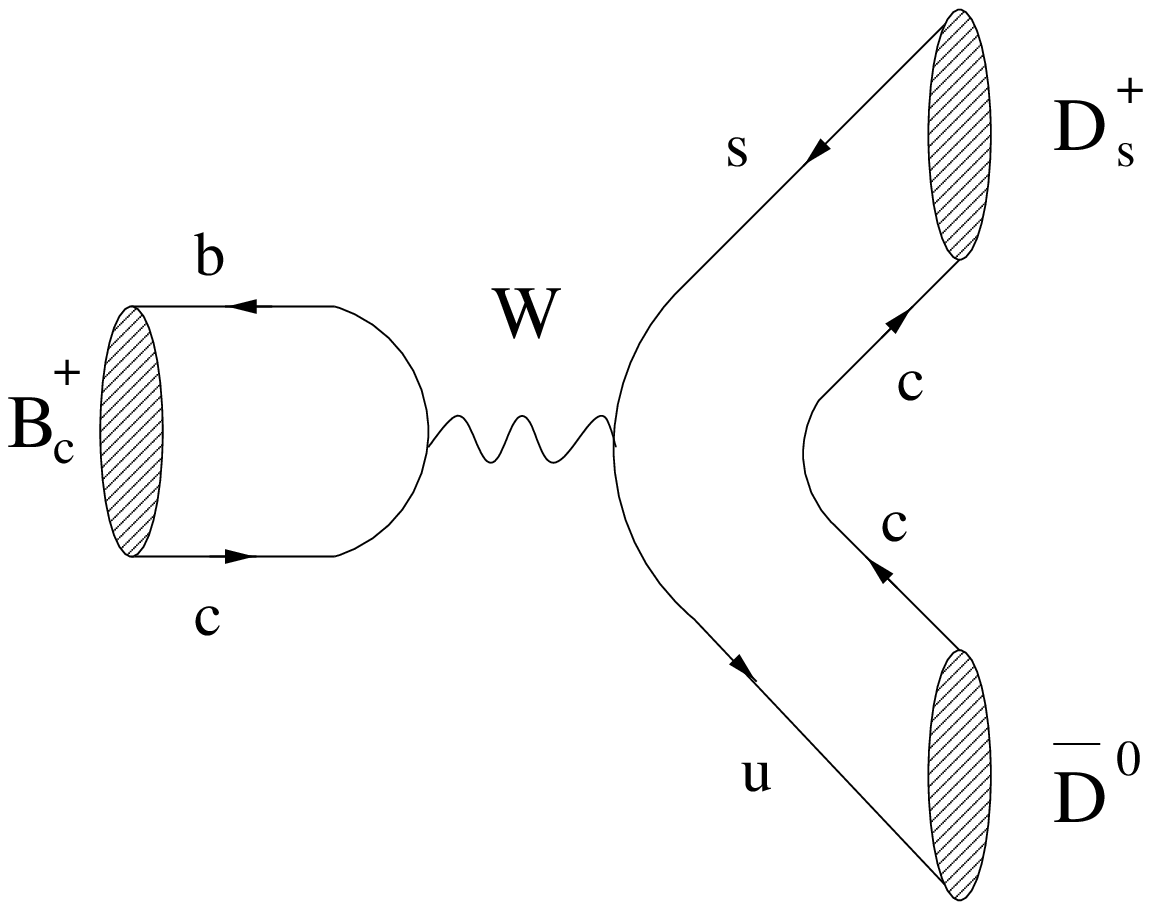}
\end{center}
\vspace*{-0.4truecm}
\caption{Feynman diagrams contributing to the decay 
$B^+_c\to D_s^+\overline{D^0}$.}\label{fig:bcdd}
\end{figure}

With the advent of hadronic $b$ facilities, it becomes possible to produce
both $B_s$- and $B_c$-mesons in large numbers. One might therefore ask to 
what extent these particles are interesting for testing the Standard Model
and, in particular, for studies of CP violation. In the case of $B_s$ decays,
there are already promising strategies \cite{LHC}. Despite the early studies 
\cite{masetti} of CP violation in non-leptonic $B_c$ decays and other more 
recent work \cite{BB}--\cite{fulvia}, no particular attention to the $B_c$
system has emerged so far. 

In this paper, we show that $B_c$-mesons could indeed play an 
important role for the exploration of CP violation. In particular, 
the $B_c$ counterpart of the $B^\pm\to K^\pm D$ triangle approach proposed 
in \cite{gw} could be well suited to extract the CKM angle $\gamma$. 
The corresponding $B_c$ decays are 
$B_c^\pm\to D^\pm_s\{D^0,\overline{D^0},D^0_+\}$, where the CP 
eigenstate $D^0_+$ introduced in (\ref{D-CP}) allows us to write the 
following amplitude relations:
\begin{eqnarray}
\sqrt{2}A(B_c^+\to D_s^{+}D^0_+)&=&
A(B^+_c\to D_s^+ D^0)+ A(B^+_c\to D_s^+ \overline{D^0})\label{rel}\\
\sqrt{2}A(B_c^-\to D_s^- D^0_+)&=&A(B^-_c\to D_s^- \overline{D^0})+  
A(B^-_c\to D_s^- D^0).\label{CP-rel}
\end{eqnarray}
The quark diagrams for these decays are shown in Figs.~\ref{fig:BcDD} and 
\ref{fig:bcdd}, where our main point can be seen: the amplitude with the 
rather small CKM matrix element $V_{ub}$ is not colour suppressed, while 
the larger element $V_{cb}$ comes with a colour-suppression factor. 
Therefore, the two amplitudes are similar in size!\, In contrast to this 
favourable situation, in the decays $B^{\pm}\to K^{\pm} 
\{D^0,\overline{D^0},D^0_+\}$, the matrix element $V_{ub}$ comes with 
the colour suppression factor, resulting in a very stretched triangle, 
while in the decays $B_d\to K^{*0} \{D^0,\overline{D^0},D^0_+\}$, all 
amplitudes are colour suppressed.

\begin{figure}
\begin{center}
\begin{picture}(320,150)(0,0)
\Line(50,10)(208,10) \ArrowLine(208,10)(210,10)
\DashLine(50,10)(290,130){6}\ArrowLine(288,129)(290,130)
\DashLine(210,10)(290,130){6}\ArrowLine(289,129)(290,130)
\Line(50,10)(130,130)\ArrowLine(128.3,128)(129,129)
\Line(210,10)(130,130)\ArrowLine(131.5,128)(131,129)
\Text(130,2)[t]{$A(B^+_c\to D_s^+\overline{D^0})=A(B^-_c\to D^-_sD^0)$}
\Text(82,80)[r]{$\sqrt{2}\,A(B^+_c\to D_s^+ D_+^0)$}
\Text(255,60)[l]{$A(B^-_c\to D^-_s\overline{D^0})$}
\Text(310,140)[br]{$\sqrt{2}\,A(B^-_c\to D_s^- D_+^0)$}
\Line(252,138)(262,118)\ArrowLine(261,120)(262,118)
\Text(95,140)[lb]{$A(B^+_c\to D^+_sD^0)$}
\Line(150,138)(145,110)\ArrowLine(145.5,113)(145,110)
\Text(210,29)[c]{$2\gamma$}\CArc(210,10)(33,57,123)
\ArrowLine(145.5,113)(145,110)
\end{picture}
\end{center}
\caption{The extraction of $\gamma$ from 
$B_c^\pm\to D^\pm_s\{D^0,\overline{D^0},D^0_+\}$ decays.}\label{fig:triangles}
\end{figure}

Taking into account that $B^+_c\to D_s^+ \overline{D^0}$ and 
$B^+_c\to D_s^+ D^0$ receive only contributions from tree-diagram-like
topologies because of the particular flavour structure of the underlying 
quark-decay processes, and that only the $\overline{b}\to\overline{u}$
transitions in Fig.~\ref{fig:BcDD} involve a CP-violating weak phase 
($\gamma$) in the Wolfenstein parametrization of the CKM matrix \cite{wolf}, 
we obtain
\begin{eqnarray}
A(B^+_c\to D_s^+ \overline{D^0})&=&A(B^-_c\to D_s^- D^0)\label{ampl1}\\
A(B^+_c\to D_s^+ D^0)&=&e^{i2\gamma}
A(B^-_c\to D_s^- \overline{D^0}).\label{gamma-rel}
\end{eqnarray}
Whereas (\ref{ampl1}) allows us to fix the relative orientation of the
two triangles described by the amplitude relations (\ref{rel}) and
(\ref{CP-rel}), (\ref{gamma-rel}) allows us to determine the CKM angle
$\gamma$, as is illustrated in Fig.~\ref{fig:triangles}. Since 
(\ref{ampl1})--(\ref{gamma-rel}) are exact in the Standard Model,
this is theoretically clean. The method is
completely analogous to the $B^\pm\to K^\pm D$ strategy \cite{gw}. 
However, as we have already noted, the advantage of the $B_c$ decays 
is that all sides of the triangles in Fig.~\ref{fig:triangles} are 
expected to be of comparable length:
\begin{equation}\label{Bc-length}
\left|\frac{A(B^+_c\to D_s^+ D^0)}{A(B^+_c\to D_s^+ \overline{D^0})}\right|=
\left|\frac{A(B^-_c\to D_s^- \overline{D^0}}{A(B^-_c\to D_s^- D^0)}\right|=
\left|\frac{R_b\left(T_c+C_c\right)}{\tilde C_c+\tilde A_c}\right|=
{\cal O}(1).
\end{equation}
Here $T_c$ and $C_c$ denote the colour-allowed and colour-suppressed
topologies in Fig.~\ref{fig:BcDD}, $\tilde C_c$ and $\tilde A_c$ describe
the colour-suppressed and annihilation topologies in Fig.~\ref{fig:bcdd},
and
\begin{equation}\label{Rb-def}
R_b\equiv\frac{1}{\lambda}\left(1-\frac{\lambda^2}{2}\right)
\left|\frac{V_{ub}}{V_{cb}}\right|=0.41\pm0.07,
\quad\mbox{with}\quad \lambda\equiv|V_{us}|=0.22.
\end{equation}
In contrast, the corresponding ratio for $B^\pm\to K^\pm D$
\cite{gw} is 
\begin{equation}\label{gw-ratio}
\left|\frac{A(B^+\to K^+ D^0)}{A(B^+\to K^+ \overline{D^0})}\right|=
\left|\frac{A(B^-\to K^- \overline{D^0}}{A(B^-\to K^- D^0)}\right|=
\left|\frac{R_b\left(\tilde C_u+\tilde A_u\right)}{T_u+C_u}\right|=
{\cal O}(0.1),
\end{equation}
resulting in the unfortunate situation, where the sides of the amplitude 
triangles involving $\gamma$ are strongly suppressed with respect to the 
remaining ones. A similar situation arises in the decays
$B_c^\pm\to D^\pm\{D^0,\overline{D^0},D^0_+\}$, obtained 
from the $B_c^\pm\to D^\pm_s\{D^0,\overline{D^0},D^0_+\}$ channels by 
interchanging all down and strange quarks ($U$-spin). 
These modes satisfy the amplitude relations
\begin{eqnarray}
\sqrt{2}A(B_c^+\to D^{+}D^0_+)&=&A(B^+_c\to D^{+}D^0)+
A(B^+_c\to D^+\overline{D^0})\\
\sqrt{2}A(B_c^-\to D^-D^0_+)&=&
A(B^-_c\to D^-\overline{D^0})+ A(B^-_c\to D^- D^0),
\end{eqnarray}
as well as
\begin{equation}
A(B^+_c\to D^+ \overline{D^0})=A(B^-_c\to D^- D^0),\quad
A(B^+_c\to D^+ D^0)=e^{i2\gamma}A(B^-_c\to D^- \overline{D^0}).
\end{equation}
Because of CKM factors different from the $B_c^\pm\to D_s^\pm D$ case, 
we obtain
\begin{equation}\label{BcU-ratio}
\left|\frac{A(B^+_c\to D^+ D^0)}{A(B^+_c\to D^+ \overline{D^0})}\right|=
\left|\frac{A(B^-_c\to D^- \overline{D^0}}{A(B^-_c\to D^- D^0)}\right|=
\lambda^2\left|\frac{R_b\left(T_c'+C_c'\right)}{\tilde C_c'+\tilde A_c'}
\right|={\cal O}(0.1),
\end{equation}
and arrive at triangles of the same shape as in the $B^\pm\to K^\pm D$ 
approach. The decays $B_d\to K^{\ast 0}D$ \cite {dun}, whose amplitudes 
are all colour suppressed and proportional to $\lambda^3(R_b)$, obviously 
have no analogue in the $B_c$ system.

As was pointed out in \cite{ads}, the small amplitude ratio (\ref{gw-ratio}) 
leads to another experimental problem: if the $D^0$-meson of the suppressed 
decay $B^+\to K^+D^0$ is tagged through the Cabibbo-favoured mode 
$D^0\to \pi^+K^-$, there are large interference effects of ${\cal O}(1)$ 
with the colour-allowed mode $B^+\to K^+\overline{D^0}[\to\pi^+K^-]$, 
where the decay of the $\overline{D^0}$-meson is doubly Cabibbo-suppressed;
indeed, all hadronic tags of the $D^0$ are affected in 
a similar way. In order to overcome these problems, it was proposed in 
\cite{ads} to use the decay chains
\begin{equation}
B^+\to K^+ D^0 \,[\to f_i],\quad B^+\to K^+\overline{D^0}\,[\to f_i], 
\end{equation}
where $f_i$ denotes doubly Cabibbo-suppressed (Cabibbo-favoured) non-CP 
modes of the $\overline{D^0}$ ($D^0$), for instance, $f_i=\pi^+K^-$ or 
$\pi^+\pi^0K^-$. If two different final states $f_i$ are considered, 
$\gamma$ can be extracted. Advantages and problems of this approach 
are discussed in Ref.~\cite{babar}. 

Because of (\ref{BcU-ratio}), it is obvious that the $B_c^\pm\to D^\pm D$
strategy is affected by similar interference problems, i.e.\ we expect
amplitudes of the same order of magnitude for the decay chains
$B^+_c\to D^+D^0[\to\pi^+K^-]$ and $B^+_c\to D^+\overline{D^0}[\to\pi^+K^-]$.
In order to extract $\gamma$, we could employ the same idea as in 
\cite{ads}. However, in the case of the $B_c$ system, an alternative is
provided by the follwing $U$-spin relations:
\begin{eqnarray}
A(B^+_c\to D^+D^0)&=&-\lambda/(1-\lambda^2/2)\, 
A(B^+_c\to D^+_sD^0)\label{U-spin1}\\
A(B^+_c\to D^+\overline{D^0})&=&(1-\lambda^2/2)/\lambda\, 
A(B^+_c\to D^+_s\overline{D^0}).\label{U-spin2}
\end{eqnarray}
Since the decay amplitudes on the right-hand sides of these equations are
of the same order of magnitude, as we have seen in (\ref{Bc-length}), the
interference effects due to $D^0,\overline{D^0}\to \pi^\pm K^\mp$ are
practically unimportant in their measurement and in the associated 
$B_c^\pm\to D^\pm_s\{D^0,\overline{D^0},D^0_+\}$ strategy to determine
$\gamma$. Consequently, this is the preferred $B_c$ approach to extract 
$\gamma$. \mbox{Nevertheless}, the Cabibbo-enhanced decay 
$B^+_c\to D^+\overline{D^0}$ plays an important role to increase the 
statistics for the measurement of the basis of the triangles shown in 
Fig.~\ref{fig:triangles} with the help of (\ref{U-spin2}). Needless to 
note, similar strategies are provided, if the $D^\pm_{(s)}$-mesons and  
are replaced by $D^{\ast\pm}_{(s)}$-mesons or higher resonances, which 
may have advantages for certain detector configurations \cite{scoop}. 

At LHC, one expects a huge number of $B_c$-mesons, about $10^{10}$
untriggered $B_c$\,s per year of running \cite{Bc-number}. The branching
ratios for the colour-suppressed $B_c$ decays were already estimated in
the literature, however, with conflicting results \cite{chinesen,fulvia}.
The following values seem reasonable:
\begin{eqnarray}
\mbox{BR}(B^+_c\to D_s^+ \overline{D^0})&\approx& 
10^{-5}\mbox{--}10^{-6}\label{CS}\\
\mbox{BR}(B^+_c\to D_s^+D^0)&\approx& 10^{-5}.
\end{eqnarray}
The first numbers for the colour-suppressed modes correspond to the range 
given in \cite {fulvia, chinesen}, while the second one for the 
colour-allowed channels is an estimate based on the results for decays
with a similar dynamics given in these papers. It is seen that the rates 
are indeed comparable. Moreover, we expect
\begin{equation}
\mbox{BR}(B^+_c\to D^+ \overline{D^0})\approx10^{-4}\mbox{--}10^{-5},
\end{equation}
allowing the measurement of (\ref{CS}) with the help of (\ref{U-spin2}). 
The predictions for the colour-suppressed 
$B^+_c\to D_s^{\ast+} \overline{D^0}$ and 
$B^+_c\to D^{\ast+} \overline{D^0}$ modes in \cite{chinesen} and 
\cite{fulvia} are in better agreement:
\begin{equation}
\mbox{BR}(B^+_c\to D_s^{\ast+} \overline{D^0})= 4\times 10^{-6},\quad
\mbox{BR}(B^+_c\to D^{\ast+} \overline{D^0})= 7\times 10^{-5};
\end{equation}
for the decay $B^+_c\to D_s^{\ast+}D^0$, we expect a branching ratio at 
the $10^{-5}$ level. 

The feasibility of the methods discussed above depends of course on the 
experimental situation and the relevant branching ratios. If we assume 
those of the final $D$-mesons to be $5\%$ and an overall efficiency of 
$10\%$, we arrive at around 20 events per year at LHC. This crude estimate 
indicates that the $B_c$ system may well contribute to our understanding 
of CP violation.

\vspace*{0.4truecm}

In this note, we have shown that decays of $B_c$-mesons appear to be
ideally suited for determining the angle $\gamma$ from triangle 
relations. The well-known disadvantages of this approach arising in 
$B_u$ and $B_d$ decays, namely small amplitudes, are absent. Provided 
there are no serious experimental problems related to the analysis
of the corresponding $B_c$ decays, this approach should be very 
interesting for the $B$-physics programme at future hadron colliders.

\vspace*{0.4truecm}

\noindent
D.W. thanks DESY for its hospitality. We thank U. Straumann and G. Wilkinson
for useful discussions, and M. Lusignoli and Z.-Z. Xing for bringing
Ref.\ \cite{masetti} to our attention.

\end{document}